\documentclass{aa}
\usepackage{txfonts}
\usepackage{graphicx}

\begin{document}

\title{AGB stars of the intermediate-age LMC cluster NGC 1846}
\subtitle{II. Dredge up along the AGB}

\author{T. Lebzelter\inst{1} \and M. T. Lederer\inst{1} \and S. Cristallo\inst{2} \and K.H. Hinkle\inst{3}
\and O. Straniero \inst{2} \and B. Aringer\inst{1,4}}

\institute{
Institute of Astronomy, University of Vienna, Tuerkenschanzstrasse 17, A1180 Vienna, Austria
\and
INAF, Osservatorio Astronomico di Collurania, 64100 Teramo, Italy
\and
National Optical Astronomy Observatories
\thanks{Operated by the
Association of Universities for
Research in Astronomy, under cooperative agreement with the National Science
Foundation.}, P.O. Box 26732, Tucson, AZ 85726, USA
\and
Dipartimento di Astronomia, Universit\`a di Padova, Vicolo dell'Osservatorio 3, 35122 Padova, Italy}

\date{Received / Accepted}

\abstract{}
{We investigate the change in the surface abundance of $^{12}$C during the evolution along the AGB,
aiming to constrain third dredge-up models.}
{High-resolution, near-infrared spectra of a sample of AGB stars in the LMC cluster NGC 1846
were obtained. A cluster sample ensures a high level of homogeneity with respect to age,
metallicity, and distance. The C/O ratio and the ratio of $^{12}$C/$^{13}$C were measured
and compared with our evolutionary models.}
{For the first time, we show the evolution of the C/O and $^{12}$C/$^{13}$C ratios along a cluster AGB. Our findings
allow us to check the reliability of the evolutionary models and, in particular, the efficiency of the third dredge up.
The increase in both
C/O and $^{12}$C/$^{13}$C in the observed O-rich stars is reproduced by the models well.
However, the low carbon isotopic ratios of the
two C-stars in our sample indicate the late occurrence of moderate
extra mixing.  The extra mixing affects the most luminous AGB stars
and is capable of increasing the abundance of $^{13}$C, while leaving unchanged
the C/O ratio, which has been fixed by the cumulative action of
several third dredge-up episodes. We find indications
that the F abundance also increases along the AGB, supporting an in situ production of this element.}
{}


\maketitle

\section{Introduction}
To understand the role of low and intermediate mass stars within
the cosmic matter cycle the efficiency of the third dredge up is a
critical quantity. During the short but decisive asymptotic giant
branch (AGB) phase the third dredge up is responsible for mixing
the burning products and the material produced via the s-process
to the surface from where it can be ejected into the interstellar
medium by stellar mass loss (see e.g. Busso et al.\,\cite{Busso99}
for a review). Current stellar evolution models (e.g. Straniero et al. \,\cite{stra97}; Herwig 
\cite{Herwig00}; Busso et al.\,\cite{Busso01}; Stancliffe et al.\,\cite{stancliffe04};
Straniero et al. \cite{stra06}; Karakas \& Lattanzio \cite{KL07}) 
provide quite
detailed predictions for the surface abundance changes during the
AGB phase. Different models agree qualitatively. The dependency of
dredge up on various parameters like mass and metallicity has
been explored in the models (Iben \& Renzini \cite{ib83};
Straniero et al.~\cite{SDCG03}).
Quantitative
checks of the model predictions for third dredge-up efficiency have been so
far restricted to AGB stars in the solar neighborhood
(e.g.\,Lebzelter \& Hron \cite{LH03}, Busso et
al.\,\cite{Busso01}) or to indirect methods like the study of the
progeny of AGB stars or synthetic stellar evolution (e.g. van Eck
et al.\,\cite{vaneck01}, Marigo et al.\,\cite{Marigo99}). However,
for studying ongoing nucleosynthesis and the third dredge-up itself,
observations of AGB stars are needed.

Among the material dredged to the surface, $^{12}$C is of special
interest as mixing of this element changes an oxygen-rich
star (C/O$<$1) into a carbon-rich one (C/O$>$1). This leads to
significant changes in the chemistry and, as a result, in the
atmospheric structure and the mass loss properties of the star.
The value of the C/O ratio is thus an indicator of the
nucleosynthesis and mixing processes inside the star. For a few
bright field AGB stars, measurements of the abundance of C and its
isotopes exist (Lambert et al.\,\cite{lambert86}; Harris et
al.\,\cite{harris87}; Smith \& Lambert \cite{SL90}). Values for
C/O range between 0.25 and 1.6 (see Fig.\,9 of Smith \& Lambert
\cite{SL90}).

Comparing the findings from field stars with nucleosynthesis and
mixing models is hampered by the rather large uncertainty in
luminosity and mass, two very critical parameters for the models.
Clusters of stars offer an excellent possibility to investigate
several questions of stellar astronomy. They provide samples of
stars that are homogeneous in age and, in general, also in
metallicity located at the same distance. Thus it is possible to
determine their evolutionary status and their mass more accurately
than for field stars. This provides an important advantage for
comparison with models of stellar atmospheres and evolution.

To investigate highly evolved stars on the AGB, globular clusters
seem the best choice due to the number of potential
targets they include and due to the age range in which they are
found. Globular clusters of the Milky Way are not good candidates when
studying the effect of the third dredge up. They are, indeed, too old
so that their present generation of AGB stars should have a very low envelope mass,
too low for the occurrence of a substantial dredge up.
The Magellanic Clouds, however, contain a population of
intermediate age clusters (e.g. Girardi et al.\,\cite{girardi95})
with AGB stars in the mass range 1.5 to 2\,$M_{\sun}$.

In this paper we present measurements of the C/O ratio and the
isotopic  ratio $^{12}$C/$^{13}$C for a sample of AGB stars in the
cluster NGC 1846. In a previous paper (Lebzelter \& Wood
\cite{LW07}), the variability of the AGB stars in this
cluster was discussed. We refer to this paper for a recent
overview of the literature on this cluster and will only repeat a
short summary here, listing the values for the global parameters
used in the present paper.

NGC 1846 is an intermediate age cluster belonging to the LMC. It
has  a metallicity of [Fe/H]$=-$0.49 (Grocholski et
al.\,\cite{grocho06}). Lebzelter \& Wood (\cite{LW07}) have determined
the mass of the AGB stars to be close to 1.8\,$M_{\sun}$ resulting
in a cluster age of 1.4$\times$10$^{9}$ years. A number of AGB
stars have been identified by Lloyd Evans (\cite{LE80}), and we
use his naming convention throughout this paper. Near
infrared measurements and estimates of the bolometric magnitudes
of the AGB stars have been presented by Frogel et
al.\,(\cite{Frogel90} and references therein). No stars with a
high mid-infrared excess, an indicator of a very high mass loss
rate, have been found (Tanab\'{e} et al.\,\cite{T98}, Lebzelter \&
Wood \cite{LW07}).

The cluster velocity has been determined by various authors in the
past. Schommer et al.\,(\cite{SOSH92})
report a cluster velocity of 240\,km\,s$^{-1}$ derived from
measurements of two individual stars (231 and 248\,km\,s$^{-1}$,
respectively). This value coincides with the result by
Freeman et al. (\cite{FIO83}) and is in good agreement with the
most recent study of individual stars in this cluster by
Grocholski et al.\,(\cite{grocho06}), who find a mean velocity of
235\,km\,s$^{-1}$ from their 17 cluster members.

\section{Observations and data reduction}

High-resolution, near-infrared spectra of 12 AGB stars in NGC 1846
from the list of Lloyd Evans (\cite{LE80}) were obtained with
the Phoenix spectrograph (Hinkle et al.\,\cite{phoenix98}) at
Gemini South during six half-nights in December 2005. We observed all 12
targets in the $K$ band between 23620 and
23700\,{\AA}; for 10 of them, spectra were also taken in the $H$
band between 15540 and 15587\,{\AA}. Spectral resolution was set
to approximately 50000. For most of the targets we achieved an S/N
ratio of 65 or better. We obained two to three observations per target and
setting on different locations along the slit to
allow for background subtraction. Total integration times between
45 and 90 minutes per target and wavelength range were used. The
spectra of each star were extracted and co-added using IRAF.

Wavelength calibration in the $K$ band was done using telluric
lines in the  spectrum of an early type star that was
also obtained. That spectrum was then used to correct the
telluric features in the $K$ band spectra of the program stars
using the IRAF task {\rm telluric}. The wavelength range in the
$H$ band was selected to be almost free of telluric lines to allow
for a better comparison with synthetic spectra. Thus we used
stellar OH lines present in the spectra of the O-rich sample stars
for the wavelength calibration in the $H$ band. The C-star spectra
were then calibrated with the solution derived from the O-rich
stars.

\section{Data analysis}
\subsection{Contents of the observed wavelength ranges}
\begin{figure}
\resizebox{\hsize}{!}{\includegraphics{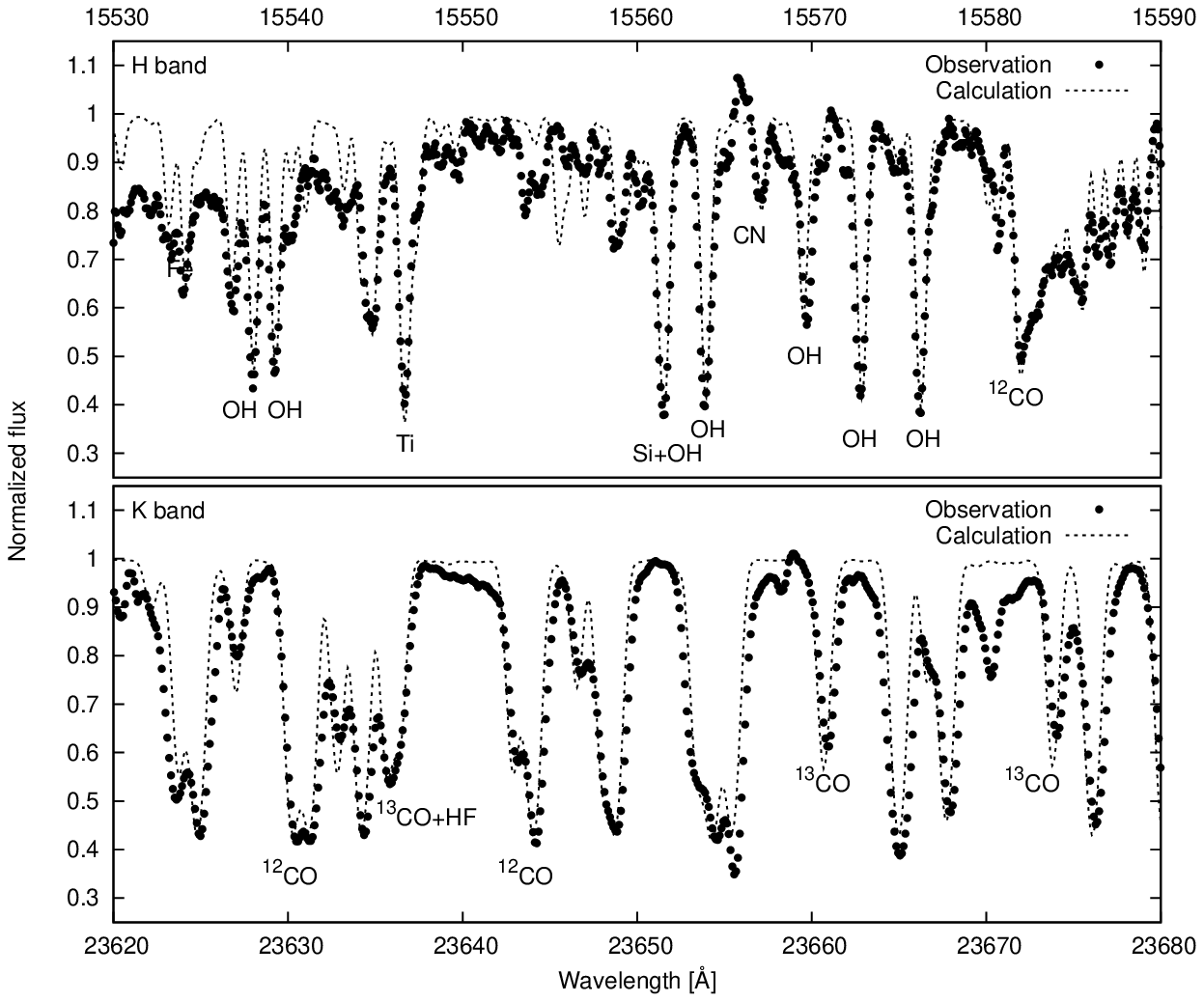}}
\caption{Observed spectral ranges. Various features are
identified. A fit to the spectrum with $T_{\rm eff}$=3550\,K,
log\,g=0.25, C/O=0.30 and a carbon isotopic ratio of 22 is given.
In this fit, the F abundance has been reduced (see text). Due to
the way the wavelength calibration was done (see text) the $H$
band spectrum is at rest wavelength while the $K$ band spectrum is
shifted in wavelength according to the stellar velocity.}
\label{specrange}
\end{figure}

The observed spectral ranges in the $H$ and $K$ bands are shown in
Fig.\,\ref{specrange} for the star LE8, i.e. for a representative
of the O-rich case. Several spectral features are identified. The
$H$ band range was selected to include a CO band head (3-0) and
several lines of OH. A few atomic lines can also be seen. Finally
a number of CN lines are present within the observed wavelength
range.

In the C-rich case, the $H$-band spectra are completely dominated
by  lines of CN and C$_{2}$. The CO 3-0 band head is present, but
its long wavelength side is strongly affected by neighboring
features of the above-mentioned molecules. Lines of CO and CN are found
in the observed wavelength range from the $K$ band.

\subsection{Synthetic spectra}
The synthetic spectra are based on model atmospheres that were
calculated with the COMARCS code, a modified version of the
MARCS code (Gustafsson et al. \cite{MARCS75}, J{\o}rgensen et
al.~\cite{MARCS92}). 
The temperature and, accordingly, the pressure stratification of the models are 
derived under the assumption of a spherical configuration in hydrostatic and 
local thermodynamic equilibrium (LTE). The LTE implies chemical equilibrium that
is evaluated in both the model calculation procedure and spectral synthesis, 
in either case taking all major opacity contributors into account such as H$_2$O, 
TiO, CO, and CN in the oxygen-rich case or HCN, C$_2$H$_2$, C$_2$, etc. in the 
carbon-rich case. For stars showing only small amplitude variability, LTE turns out to be a 
valid approximation, while potential uncertainties in the determination of abundances 
are due to a deviation from hydrostatic equilibrium rather than to non-LTE effects.
COMARCS makes use of updated atomic and molecular
opacities calculated with COMA (Aringer \cite{Ari2000}). In 
calculating the model spectra, spherical radiative transfer
routines originating from Windsteig et al.~(\cite{Windsteig97})
were used. The atomic line data were taken from VALD (Kupka et
al.~\cite{VALD}), whereas the molecular data were compiled from
several sources (see Cristallo et al.~\cite{cri07} for an overview and Lederer \& Aringer~\cite{LA07} for more recently added line lists).
Model atmospheres and synthetic spectra that were calculated following the method
described above have been shown to describe the spectra of cool
giant stars appropriately (e.\,g. Loidl et al. \cite{Loidl01},
Aringer et al. \cite{Ari2002}).

We want to point out that we calculated
the model structure and the synthetic spectra in a fully
consistent way with respect to spherical geometry and the atomic and
molecular input data. Each atmospheric model is determined by six parameters:
mass,  metallicity, effective temperature, surface gravity, C/O
ratio, and microturbulent velocity. From the full set of
parameters we kept the mass ($M=1.5\,M_\odot$) and the metallicity
([M/H]=$-0.4$) fixed (cf. next section for more details and
references). The effective temperature was varied in steps of $50\,{\rm K}$, and
the values for $T_{\rm eff}$ range from $2600\,\rm{K}$ to
$3850\,\rm{K}$ in our model set. In terms of the surface gravity,
we chose a stepsize of $0.25$ in logarithmic scale, i.\,e.
$\log\,(g [cm\,s^{-2}])$, ranging from $-0.50$ to $+0.50$. The
microturbulent velocity in the model calculations was set to
$\xi=2.5\,km\,s^{-1}$ apart from the model for the star LE13 where
we adopted a value of $\xi=3.5\,km\,s^{-1}$ (see below for details).
The solar abundances were taken from Grevesse \& Noels
(\cite{GN93}) whereby an inherent value of C/O = $0.48$ emerges.
As for the variation in the C/O ratio, we altered the abundance of carbon
to arrive at the desired value and left the other abundances untouched. We
changed the C/O ratio in steps of $0.1$ in general, but also
reduced the stepsize to $0.05$ when necessary for the fitting
procedure (see next section).

The wavelength range covered in the calculation of the synthetic
spectra  was $6400$-$6450\,{\rm cm^{-1}}$ ($15504$-$15625$ \AA) in
the H band and $4215$-$4270\,{\rm cm^{-1}}$ ($23420$-$23725$ \AA)
in the K band. The synthetic spectra were calculated with a
resolution of $300 000$ and then convolved with a Gaussian 
to get a resolution of $50 000$, which is about the value attained
in our observed spectra. Subsequently, the macroturbulent velocity
was accounted for by another convolution with a Gaussian. In the
course of our analysis we also varied the carbon isotopic ratio
$^{12}$C/$^{13}$C. The solar value as given in Anders \& Grevesse
(\cite{AG89}) is about $89.9$. The variation in the isotopic ratio
has de facto no effect on the structure of the model atmosphere
and thus was only considered in the spectral synthesis calculations.
In the next section we give details in the variation of the 
above-mentioned parameters for the individual targets.

\subsection{Abundance determination}
To determine elemental abundances we used a two-step process. As a
starting point we used $T_{\rm eff}$ and $L$ values derived from
broad-band, near-infrared photometry as described in Lebzelter \&
Wood (\cite{LW07}). Assuming a stellar mass of 1.5\,M$_{\sun}$, we
derived a starting value for the stellar radius and the surface
gravity. Around these parameters we calculated a small grid of
synthetic $H$ band
spectra and selected the best fit with the observed spectrum both by visual
inspection and by a $\chi^{2}$ approach. Several relative line
strengths could be used to estimate the correctness of the fit.
We varied $T_{\rm eff}$, log\,g, and C/O ratio; among these three
parameters, changes in log ${g}$ had the weakest effect. Thus a small
uncertainty in the mass or radius estimate of the star only has a
minor influence on the derived abundances\footnote{We thus do also
not have to bother about the small difference in mass between the one
derived in Lebzelter \& Wood (\cite{LW07}) and the value assumed
for calculating the synthetic spectra.}. In the O-rich case,
changes in $T_{\rm eff}$ have the strongest effect on the OH/Fe
blend at 15570\,{\AA}, but an increase in temperature leads to a
general increase in the strength of all features. An increase in
C/O makes the CO band head stronger and all the
OH lines weaker at the same time. Additionally such a change leads to more
prominent lines of CN. In the chosen
wavelength range, the model can be fitted to a well-defined
pseudocontinuum. All these facts allow us to make good fits of the spectra
in the O-rich case as
illustrated in Fig.\,\ref{specrange} and to derive a C/O ratio
from that. To achieve a good match of the lines, the macroturbulence
was set to 3\,km\,s$^{-1}$. The influence of other
values was tested. For the fit of LE16 and LE13, a
macroturbulent velocity of 4 and 7\,km\,s$^{-1}$, respectively, was
necessary to account for the broader lines observed in these
stars. For LE13 we also increased the microturbulence to
3.5\,km\,s$^{-1}$ as mentioned before. We note that it was not our aim to measure the
metallicity of the stars from our spectra because the selected
wavelength ranges include only a small number of weak or strongly
blended metal lines. The chosen metallicity of [Fe/H]$=-$0.4 for
our models is close to the literature value and the metallic lines
in the synthetic spectra are in reasonable agreement with the
observed spectra. The fitting of the C-stars turned out to be much
more complicated (see below).

In a second step, the results from the $H$ band spectrum were used
as  starting parameters for fitting the $K$ band spectrum. The
latter is -- especially for the O-rich case -- rather insensitive to
changes in $T_{\rm eff}$, log ${g}$, and C/O ratio. Thus these
parameters cannot be determined from the $K$ band spectra alone.
The chosen value of the macroturbulence could be confirmed. The main
purpose for obtaining and analyzing the $K$ band spectra was to
determine the $^{12}$C/$^{13}$C isotopic ratio, which with
the C/O ratio is also a very reliable indicator of third
dredge up. For this the O-rich spectra offer two (almost)
unblended lines of $^{13}$CO (2-0 R18 and 2-0 R19) and two blends
including contributions from $^{13}$CO lines. The best fit was
selected by minimizing the squared difference in the spectral
range around the two unblended $^{13}$CO lines. Again, fitting of
the C-stars was more problematic and is briefly discussed
separately below.

Uncertainties of the abundances were estimated from the range of
parameters of the model spectra that still gave a good fit. For
the C/O ratio, the $T_{\rm eff}$, and the log ${g}$, value this was done
from the fit of the $H$-band spectrum, for $^{12}$C/$^{13}$C from
the fit of the $K$-band spectrum. Because the carbon
isotopic ratio has a negligible effect on the fit of the $H$-band
spectrum, as verified by test calculations, the fitting procedure
was not repeated with the determined $^{12}$C/$^{13}$C ratio.

The $K$ band spectra also include a blend of the $^{13}$CO 2-0 R21
line with an HF line. When comparing this line blend with the synthetic spectrum
using the isotopic ratio derived from the two unblended $^{13}$CO lines,
it turned out that this blend could not be fitted properly. An
improvement in the fit can be either achieved by a change in
$T_{\rm eff}$ or by a modification of the F abundance. The first
one was only possible in a very limited range due to the
constraints set by the fit of the $H$ band spectrum. Thus we
modified the abundance of F until we achieved a good fit. This was
only done for the O-rich stars because in the C-stars the blend is
additionally affected by a nearby CN line.

The lowest excitation lines of CO in our $K$ band spectra could
not be  fitted well, especially in the case of the more luminous
O-rich stars in our sample. An explanation for this could not be
found; however, it may indicate a problem of the model structure in
the outermost layers of the star, which may reflect dynamic
effects or an outflow of material expected for variable stars.

\subsection{Carbon stars}
When fitting the C-rich stars, we were confronted with two major
problems.  The first one concerns the uncertainties of the line
lists for the C-bearing molecules. The line lists of CN and C$_{2}$, both
critical for fitting the $H$-band spectrum, produced a number of
features in the synthetic spectra that were obviously either at
the wrong wavelength or of the wrong intensity. The CN list could
be improved by replacing the wavelengths in the line list with the
observed ones taken from Davis et al. (\cite{cnlist05}). This was
possible for a number of strong lines, while the weak lines
typically had no counterpart in the observed catalog. After that,
a few CN lines with wrong line strengths still showed up in the
synthetic spectrum. One of these lines, located at
6417.405\,cm$^{-1}$, was removed by hand from the CN list for the
later calculation. This line is also not in the catalog of
observed lines.

No observed reference is available for the C$_{2}$ line list. Thus
correcting this list is only possible by visual comparison of our
spectra of C-stars in NGC 1846 with synthetic spectra. Within the
studied range in the $H$-band, 5 lines were removed completely and
9 were shifted in wavelength to fit the observed
spectra better. In this wavelength range, lines of the two molecules
including a $^{13}$C atom (i.e. $^{13}$C$^{12}$C and
$^{13}$C$^{14}$N) play an important role for some of the blends.
It turned out to be impossible to check the quality of these line
data or to improve the line lists in that case because determining the
carbon isotopic ratio was one of the aims of our project. We 
come back to this point in \ref{isotopic}.

The second major problem was the absence of reliable `continuum'
points in the C-star spectra. The main effect of decreasing
$T_{\rm eff}$ or increasing the C/O ratio is the amount of
depression of the (pseudo)continuum by a large number of weak and
densely spaced lines over the whole spectral range. Thus the
overall observed line depths can be fitted with a variety of
combinations of these two parameters simply using different
scaling factors between the observed and the calculated spectrum.
At first glance, a possible point for the pseudocontinuum seems
to be present close to 15571\,{\AA}. However, closer inspection
reveals an absorption feature at that position in the observed
spectrum that is missing in the synthetic spectrum. Therefore we
think that this point is also not useable as a reference point.
Thus we are limited to look for the comparably weak effect of
using the ratios of features that seem to rely on one or the other
of the two parameters. This, however, is hampered both by the
uncertain line lists as discussed above and by the fact that a
deeper CN or C$_{2}$ line can either be achieved by an increase in
C/O or a decrease in temperature. For determining the C/O ratio, we
thus focused on the CO 3-0 band head that seems to be the only
feature decoupled from this trend, but a larger uncertainty of the
result compared to the O-rich case was unavoidable. The CO band
head weakens relative to its neighboring features as C/O is
increased, and it is almost insensitive to temperature changes.

Fitting the $K$-band is eased by the lack of strong C$_{2}$ lines, but as
in the O-rich case, the depth of the various spectral features is
rather insensitive to changes in the global parameters.
Unfortunately, one of the two unblended $^{13}$CO lines from the
O-rich case is heavily blended by CN in the C-rich case. That
blending CN line seems to be real, as the shape of the two lines
also differs in the observations, but the strength of the CN line
is wrong in the synthetic spectrum (independent of changes in the
global parameters). Thus this line could not be used and the
determination of the isotopic ratio could be based only on one
line with some additional indications from one blended feature in
the C-rich case.

\subsection{Evolutionary models}
The theoretical models we use in the discussion of the observational results
have been obtained by means of the FRANEC
stellar evolutionary code (Chieffi et al.\,\cite{chi98}). A detailed description of the
input physics can be found in Straniero et al.\,(\cite{stra06}) and Cristallo et
al.\,(\cite{cri07}). We calculated evolutionary sequences for various initial masses
and compositions. In particular, models labeled as {\it scaled solar}, were obtained by scaling
the solar abundances derived by Lodders\,(\cite{lo03}) of a factor Z/Z$_\odot$, 
where Z is the assumed metallicity of the model.
Then, {\it $\alpha$-enhanced} models were obtained by increasing the initial abundance 
of O, Ne, Mg, Si, S, Ca, and Ti by a common factor with respect to the corresponding scaled solar values
(see Piersanti et al.\,\cite{pier07} for further details).

\section{Results}
\subsection{Cluster membership}
From the K-band spectra, we measured radial velocities of the
target  stars from individual lines of CO. Results
are presented in Table\,\ref{velos}. For LE8, Olszewski et
al.\,(\cite{OSSH91}) give a velocity of 248$\pm$5\,km\,s$^{-1}$, in
excellent agreement with our value. The mean velocity of all
investigated AGB stars in NGC 1846 is 248$\pm$5.5\,km\,s$^{-1}$,
which is in good agreement with the published values for the
cluster. The only star with a larger deviation from the mean
velocity is LE17.

\begin{table}
\caption{Radial velocities of NGC 1846 AGB stars. Concerning the membership of LE17
please see the discussion in the text.}
\label{velos}
\centering
\begin{tabular}{l r | l r}
\hline \hline
Star & rv [km/s] & Star & rv [km/s]\\
\hline
LE1 & 251$\pm$2 & LE11 & 244$\pm$3\\
LE2 & 248$\pm$2 & LE12 & 247$\pm$2\\
LE3 & 248$\pm$3 & LE13 & 242$\pm$2\\
LE6 & 247$\pm$2 & LE17 & 262$\pm$1\\
LE8 & 250$\pm$1 & H39 & 242$\pm$1\\
LE9 & 245$\pm$1 & & \\
\hline
\end{tabular}
\end{table}

\subsection{O-rich stars}
The C/O ratio and the $^{12}$C/$^{13}$C ratio were determined
for all six O-rich stars in our sample. The C/O ratios between 0.2 and
0.65 were found with an uncertainty between $\pm$0.05 and
$\pm$0.1, and $^{12}$C/$^{13}$C isotopic ratios between 12 and 60 were found.
Temperatures of the best fit model spectra agree closely
with the $T_{\rm eff}$ values derived from near infrared
photometry (Lebzelter \& Wood \cite{LW07}). The result for each
star is listed in Table \ref{t:coetc} with the basic
parameters of the stellar models used.

\begin{table*}
\caption{Resulting abundances and model fit parameters for the sample stars. The last two
columns give the periods and luminosities taken from Lebzelter \& Wood (\cite{LW07}).}
\label{t:coetc}
\centering
\begin{tabular}{lcccccc|cc}
\hline \hline
Star & C/O & $^{12}$C/$^{13}$C & [F/Fe] & $T_{\rm eff}$ & log ${g}$ & macroturbulent & P [d] & log\,$L$/L$_{\odot}$\\
  &  &  &  &  &  & velocity & & \\
\hline
LE2 & 1.9$\pm$0.2 & 65$\pm$15 & -- & 2600 K & -0.25 & 10 & 150 / 289 & 3.835\\
LE8 & 0.3$\pm$0.05 & 20$\pm$2 & $-$0.35 & 3550 K & 0.25 & 3 & 51 & 3.719\\
LE9 & 0.2$\pm$0.05 & 13$\pm$2 & $-$0.42 & 3650 K & 0.25 & 3 & -- & 3.604\\
LE11 & 1.7$\pm$0.2 & 60$\pm$10 & -- & 2900 K & -0.25 & 10 & 227 / 1066 & 3.842\\
LE13 & 0.65$\pm$0.1 & 60$\pm$5 & +0.40 & 3600 K & 0.0 & 7 & 92 / 837 & 3.806\\
LE16 & 0.44$\pm$0.05 & 43$\pm$2 & -0.20 & 3600 K & 0.0 & 4 & 57 & 3.645\\
LE17 & 0.44$\pm$0.05 & 30$\pm$4 & +0.08 & 3700 K & 0.0 & 3 & 61 & 3.491\\
H39 & 0.2$\pm$0.05 & 12$\pm$2 & $-$0.71 & 3650 K & 0.25 & 3 & 33 & 3.538\\
\hline
\end{tabular}
\end{table*}

The C/O ratio obviously varies significantly from star to star,
indicating different amounts of enriched material dredged up to the
surface. It would be expected that an increased C/O ratio due to
the dredge up of $^{12}$C would be accompanied by an increased
isotopic abundance ratio. Indeed we find a nice correlation
between C/O and $^{12}$C/$^{13}$C for the O-rich stars as can be
seen from Table \ref{t:coetc}. The values for C/O and
$^{12}$C/$^{13}$C ratio found are in good agreement with findings
from Smith \& Lambert (\cite{SL90}) for field stars. The minimum C/O ratio
in our sample of 0.2 agrees closely with expectations from first dredge-up.

As described above, one blend including a line of HF (1-0 R7) is
covered by our $K$ band observations. After limiting the various
parameters affecting the shape and depth of this blend, namely
mainly the temperature and the carbon isotopic ratio, from other
parts of the stellar spectra we fitted the remaining difference
between the observed and the calculated blend profile by a change
in the F abundance. Due to the rather large uncertainties of the
C-star fitting we decided to derive the F abundance only for the
O-rich stars.

\begin{figure}
\resizebox{\hsize}{!}{\includegraphics{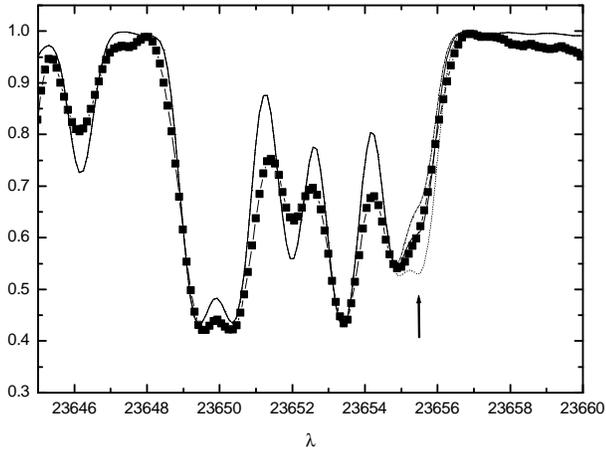}} \caption{Part of
the $K$ band spectrum of LE8. The location of  the HF line is
marked by an arrow. Beside the spectrum (line + symbol) three
models are plotted with a variable F abundance of $-$0.1 dex
(dotted), $-$0.5 dex (solid), and $-$0.7 dex (dash-dot) relative to a
solar value (-7.44) and scaled for the average cluster
metallicity, respectively.} \label{hfplot}
\end{figure}

Figure \ref{hfplot} shows as an example part of the spectrum of LE8
including the HF line (marked by an arrow). Overplotted are three
models with different F abundances. Obviously, Fluorine is
underabundant in LE8 relative to the other metals. It turns out
that the F abundance has to be changed from star to star to allow
for a reasonable fit of the observed blend. Uncertainty in
determining the F abundance comes from both the temperature and
the $^{12}$C/$^{13}$C ratio (via a $^{13}$CO line in the blend).
We made an estimate of the uncertainty by varying these two
parameters around the values determined from other parts of the
spectra and then summing in quadrature the respective uncertainties.
Taking into account this uncertainty and the fact that we
have only one HF line to derive the abundance the results of
course have to be taken with some caution. Confirmation from
other HF lines is needed to limit the possible effect of unidentified
lines on the profile of the blend and the derived abundances.

\subsection{C-rich stars}\label{isotopic}
The C/O ratio was derived for two C-rich stars, LE11 and LE2. For the
other C-rich stars, either no $H$ band spectra were available (as a
result of limited observing time) or no acceptable fit of the
spectrum could be achieved. For the latter case, the starting
influence of stellar variability on the stellar spectrum may play
some role besides the above-mentioned problems.

The derived $T_{\rm eff}$ and C/O for the C-rich star LE11
come with a much larger uncertainty. From the strength of the CO
3-0 band head, we exclude a C/O of less than 1.4. A
macroturbulent velocity of 10\,km\,s$^{-1}$ is needed to both fit
the width of the features and their strength. The temperature can
be constrained even less. If we use the value from near infrared
photometry, we get a $T_{\rm eff}$ of 2950\,K. As noted by
Lebzelter \& Wood (\cite{LW07}), this temperature is likely to be too low
as pulsation properties instead suggest a value close to 3150\,K.
We produced model spectra around both temperatures. A value of $T_{\rm eff}$=2950\,K
seems to give the better fit, but the sensitivity of the synthetic spectrum
on this parameter is not very high.

\begin{figure}
\resizebox{\hsize}{!}{\includegraphics{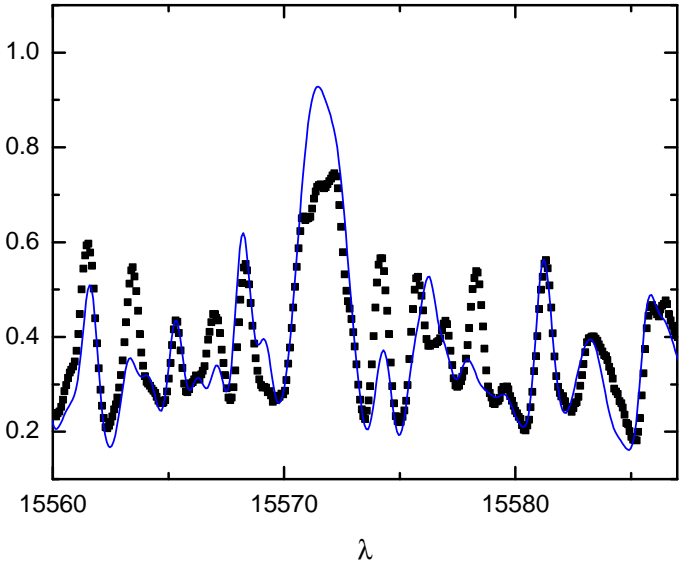}}
\caption{Spectrum of the carbon star LE11 compared to a model fit
with C/O$=$1.7 and $T_{\rm eff}$=2950\,K. See text for details.}
\label{le11fit}
\end{figure}

Concerning an upper limit of the C/O ratio an overall fit of
the spectrum of similar quality can be reached for various
combinations of C/O and $T_{\rm eff}$ with a higher temperature
requiring a higher C/O ratio. As a rule of thumb an increase in $T_{\rm eff}$ by 100 K can be 
compensated by an increase of C/O by 0.1 in the parameter region under consideration. Taking the
uncertainties of the line lists into account we cannot exclude a C/O ratio as
high as 2.0, but an upper limit of 1.9 seems reasonable due
to constraints set by the range of possible temperatures defined by the K band spectrum,
the near infrared color and the pulsational behavior, respectively.
In Fig.\,\ref{le11fit} we show a fit with C/O$=$1.7
and $T_{\rm eff}$ of 2950\,K.

With this value we derived a $^{12}$C/$^{13}$C ratio
of about 60 from the $K$ band spectrum. The derived isotopic ratio depends on
the temperature, the C/O ratio, and the selection of the point in the synthetic
spectrum to which the observed one is scaled. The C/O ratio has the least
influence and, within the constraints set by the $H$ band spectrum, its uncertainty
can be neglected. The temperature is an important parameter for $T_{\rm eff}$
above 2900\,K. Below that value, uncertainties in this parameter affect
the resulting $^{12}$C/$^{13}$C ratio less and less. The selection of the `pseudocontinuum'
is of high importance in this low temperature regime. From a comparison between
the observed spectra and synthetic spectra by varying the mentioned parameters,
we derive a possible range for the isotopic ratio of $\pm$10 around the mean value.

Additionally we were able to produce a reasonable fit for the spectra of
LE2 when using a temperature between 2600 and 2850 K, the $H$ band spectrum
indicates a C/O ratio of 1.9$\pm$0.2. The lower temperature nicely corresponds
to the one derived from the $J-K$ value (Lebzelter \& Wood \cite{LW07}).
The resulting $^{12}$C/$^{13}$C ratio is close to 65$\pm$15.

While we were not able to make a good fit of the $H$ and $K$ band
spectra of LE1, we made a qualitative comparison of LE1 and LE2.
Both wavelength ranges indicate that LE1 has a higher C/O ratio
than LE2. Unfortunately, a quantitative estimate was not possible.

\section{Discussion}
\subsection{Dredge up along the AGB}
\begin{figure}
\resizebox{\hsize}{!}{\includegraphics{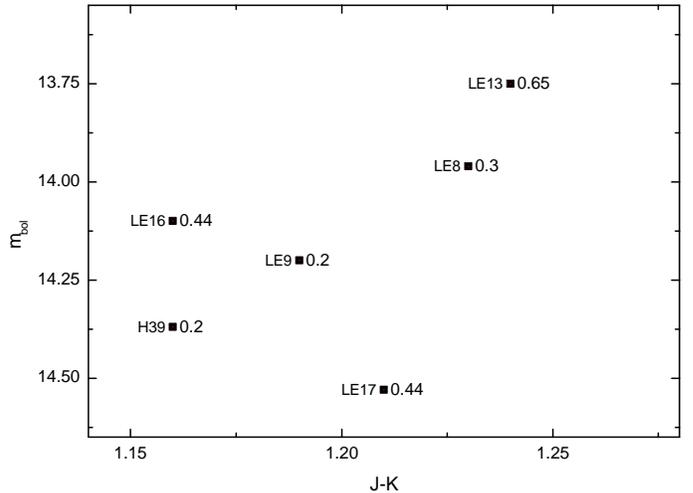}}
\caption{Color-magnitude diagram of the O-rich stars in our
sample. The C/O ratio of each data point is given.} \label{cmd1}
\end{figure}

In Fig.\,\ref{cmd1} the location of the O-rich stars of our sample
in a color-magnitude diagram are shown. The data points are
labeled with their respective C/O ratio. 
Color
and $m_{bol}$ values used here are taken from Lebzelter \& Wood
(\cite{LW07}). As shown in that paper, color variations of these
stars are very small and can be neglected in the present discussion.
The $m_{bol}$ values are derived from the near infrared magnitudes
using the transforms in Houdashelt et al.\,(\cite{Houdi00a},\cite{Houdi00b}).
For details we refer to Lebzelter \& Wood (\cite{LW07}).
The sample stars cover a significant range in
brightness along the AGB, thus allowing us to look for changes in the
C/O and $^{12}$C/$^{13}$C ratios with increasing luminosity.
%%%%changed
We can exclude that these stars belong to the RGB. Indeed, the RGB
tip observed by Lebzelter \& Wood (\cite{LW07}) is definitely
fainter than the faintest star in the sample, LE17. This
occurrence is confirmed by our evolutionary models. Actually, LE17
is only slightly brighter than the RGB tip of the 1.5 M$_\odot$
model, but its $^{12}$C/$^{13}$C ratio (30) is definitely greater
than expected for the brightest RGB stars (see
e.g. Sneden \cite{sneden}, Gratton et al \,\cite{gratton}).
%%%%%%
LE13, the brightest star in our sample, is also the brightest
O-rich AGB star known in this cluster. However, the resulting
picture is  far from clear. Indeed we find the star with the
highest C/O value at the top. But the two stars with a C/O ratio
close to 0.44 (LE16 and LE17) do not seem to fit into a steady
increase with luminosity at all. They also seem to be offset from
the other stars,
 which seem to form some sequence in the CMD.

We note here that the luminosity at which an AGB star is observed
does not necessarily have to be representative of its
evolutionary state along the AGB. During thermal pulses the
brightness can change significantly both to lower and to higher
values. It was suggested in Lebzelter \& Wood (\cite{LW07}) that
at least the atypically low luminosity of two of the carbon stars
in this cluster can be explained by a deviation from the mean
interpulse luminosity during a thermal pulse.
%%%%%changed

LE17 seems to be an outlier in various aspects. Assuming
that the star is a cluster member, its relatively low luminosity, 
coupled to its relatively
large  $J-K$, could be explained if this star is undergoing a
post-flash dip  phase (Iben \& Renzini \cite{ib83}). The expansion
powered by a thermal pulse indeed causes a significant decrease
in the luminosity and in the effective temperature. From our
models, we derive that variations up to 1 mag  and 200 K (in
M$_{bol}$ and $T_{\rm eff}$, respectively) have to be expected. Alternative
scenarios for explaining the characteristics of LE17 could be
either an unresolved binary or a long secondary period similar to the ones
already found in several AGB stars (e.g. Wood et al.~\cite{Wood04}). The observed offset in velocity of LE17
from the other AGB stars in our sample could be consistent with
both of these scenarios. The star's obvious enrichment in
$^{12}$C, however, could be best explained by the post-flash dip
scenario.

An alternative would be that LE17 is not a cluster member
but a background star. Indeed this solution cannot be ruled out.
Frogel et al. (\cite{Frogel90}) list the star as a cluster member
based on its projected distance to the cluster center. Lebzelter \& Wood
(\cite{LW07}) used the location of the star in the color-magnitude diagram
as an argument in favor of the cluster membership.
On
the other hand, the radial velocity of the star is offset from the other
AGB stars measured by almost 20\,km\,s$^{-1}$. In their work
on the velocity dispersion in LMC clusters Grocholski et al.\,(\cite{grocho06})
chose a range in velocity of $\pm$10\,km\,s$^{-1}$ around the mean cluster velocity
to separate members from non-members. According to this rule LE17 would not
be a member. Based on the present data, it is difficult to distinguish between the
two possibilities.

Concerning
the other ``anomalous" star, LE16, one could speculate that
the observed AGB stars may belong to two sequences, 
one including the stars with C/O ratios
between 0.2 and 0.3, and another one, shifted to the blue, with
the two remaining stars. We come back to this point below.
%%%%%%%%%%

As mentioned above there is a qualitative agreement in the
derived abundances between the increase in C/O and
$^{12}$C/$^{13}$C. Thus, a wrong abundance for the two deviating
stars is not very likely.
%%%%changed
Note that all the stars with C/O$>$0.2 and $^{12}$C/$^{13}$C$>$15
but LE17 are brighter than the minimum luminosity for the
occurrence of the third dredge up, as predicted by our models.
Taking the mean luminosity of the interpulse period just after the
occurrence of the first TDU episode, we find that all stars with
m$_{bol}$ brighter than 14.1 mag should be $^{12}$C enhanced. Once
again, the anomaly of LE17 could be solved if the star was in the
post-flash dip.
%%%%%%%%

\begin{figure}
\resizebox{\hsize}{!}{\includegraphics{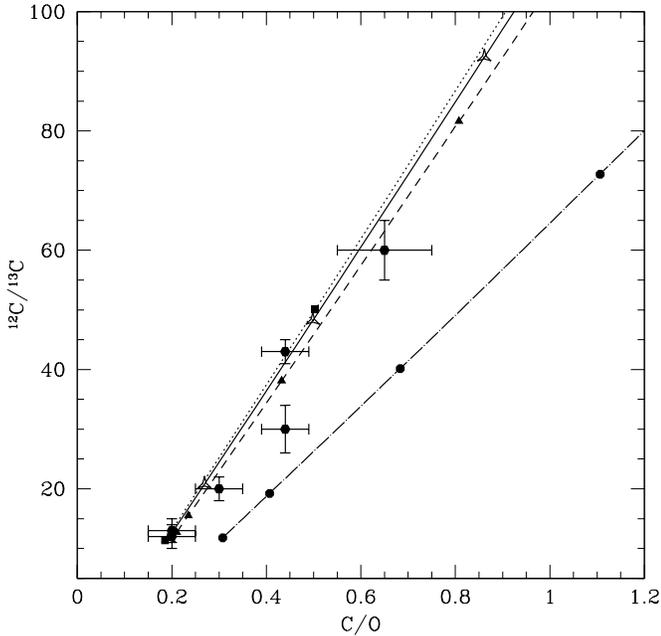}}
\caption{C/O versus $^{12}$C/$^{13}$C for the O-rich stars in our
sample. For comparison predicted values from our evolutionary
models are shown. The lines connect individual interpulse values
for four sets of model parameters. See text for details.}
\label{obsmeetmodels}
\end{figure}

 We illustrate the relation between
C/O and $^{12}$C/$^{13}$C in Fig.\,\ref{obsmeetmodels}. While
there  is some scatter, a relation between these two quantities can
clearly be seen.
%%%%%changed
In the same figure, we also plotted the expected relations from
evolutionary models. The solid line represents our reference
model: M=1.9 M$_\odot$ (1.8 at the beginning of the TP-AGB phase),
Z=0.006, [$\alpha$/Fe]=0.2, Y=0.27. The mass has been chosen
according to the pulsational mass derived by Lebzelter \& Wood
(\cite{LW07}), while the moderate enhancement of the $\alpha$
elements corresponds to the value generally claimed for the Large
Magellanic Cloud (see e.g. Hill et al.~\cite{Hill00}). 
The C/O ratio of the two less evolved stars in our sample, H39 and
LE9, supports such an assumption.
Indeed, the expected C/O ratio after the first dredge up is 0.35,
for a scaled solar composition, and 0.2 when [O/Fe]=+0.2. In this case,
the total metallicity (Z=0.006) would imply [Fe/H]=-0.5, in agreement
with the latest spectroscopic determination for NGC 1846 (Grocholski et al.
\cite{grocho06} ). To evaluate the effects of possible variations of these
model parameters, we also report the relations obtained by
changing the mass (1.5 M$_\odot$, dashed line), the total
metallicity (Z=0.003, dotted line), and the initial abundance
ratios (scaled solar instead of $\alpha$-enhanced, dot-dashed
line).
%%%%%%%%%%
%%%%%changed

It can be seen that all data points are found very close to the
theoretical relation for the reference model,  even if LE17
presents a lower carbon isotopic ratio (or too high
C/O)\footnote{Note that the abundances have been determined with a
model atmosphere without O enhancement. To see if this has any
effect on the derived ratios, we calculated a test model with O
enhanced by +0.1 and +0.2 dex, respectively. It turned out that this enhancement
only has a weak effect on the spectra which can be compensated
for by small changes in the temperature. The C/O and the
$^{12}$C/$^{13}$C ratios remain almost unchanged.}. Changes in
initial mass and metallicity have little effect, so that the
uncertainties of the present spectroscopic analysis do not allow us
to distinguish among the different relations reported in Fig. 5.
The model with a solar-scaled composition can instead be safely
excluded.
%%%%%%%%%
The boxes along the relations mark the interpulse values after
each dredge up event. Close to these  values we should have the
highest probability of finding a star. In general, the agreement is
remarkably good, with one exception (LE13).

%We would further expect to find a limited number
%of different C/O ratios and again our observations are in
%reasonable agreement with the model.

\begin{figure}
\resizebox{\hsize}{!}{\includegraphics{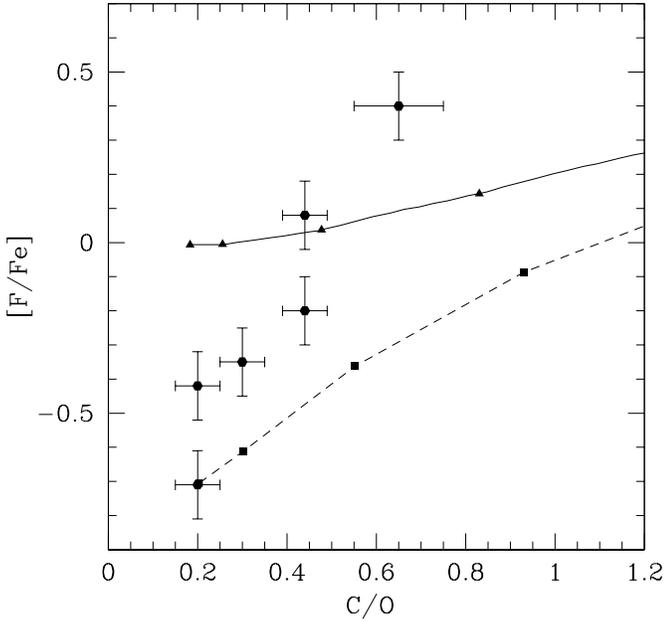}}
\caption{Fluorine abundances of our O-rich stars sample as a
function of the C/O ratio. The solid line corresponds to the
reference theoretical model, while the dashed line represents a model 
for which the initial [F/Fe] is -0.71 instead of 0.} \label{Fabun}
\end{figure}

%%%%changed
Concerning the abundance of the light element F, we observe, as
shown in Fig.\,\ref{Fabun}, a change from an F underabundance to an
F overabundance with increasing C/O (relative to the solar value
scaled to the iron abundance of the cluster). Accordingly, there is
also a similar trend with the luminosity, as found for the C/O
ratio. Measurements of HF lines are a critical source for
obtaining the F abundance in stars.
Various hypotheses on the origin of this element have been discussed in the literature
(see Schuler et al.\,\cite{SCS07} for a recent overview).
In particular, from a correlation of the
carbon abundance with the F abundance, Jorissen et al.
(\cite{jo92}) showed that nucleosynthesis in AGB stars is one
likely path for the production of this element.
Schuler et al. (\cite{SCS07}) come to a similar result from
F abundance measurements of a carbon-enhanced, metal-poor star. 
A positive correlation between F and C/O has also been acknowledged 
in a planetary nebula (Zhang \& Liu \,\cite{zhang05}).
Our data of NGC 1846 AGB stars clearly support the
production and dredge up of F during the AGB evolution. The two 
lines in the figure represents the predictions of the reference
model. The only difference between the two lines is in the initial fluorine abundance. 
We have assumed [F/Fe]=0 and [F/Fe]=-0.71 at the beginning of
 the evolutionary sequence, for the solid and the dashed lines, respectively. 
The lower value corresponds to the one of the less 
evolved AGB star in our sample (H39). 
Note that Cunha et al. (\cite{cunha03}) report similar fluorine underabundances
in LMC giants.
Our models predict a negligible variation in the surface F abundance after the first dredge up.
It should be noted that only an ``extreme'' extra mixing during the RGB may induce a sizeable 
F depletion (see e.g. Denissenkov et al. \cite{deni06}).
In contrast, during the AGB fluorine is produced by $\alpha$ captures on  $^{15}$N in
the convective zone generated by a thermal pulse and, later on, dredged up. Part of the
$^{15}$N seeds are in the ashes of the CNO burning (about 50\%), while the
remaining part is synthesized in the $^{13}$C pocket. In both cases, 
$^{15}$N is released by the
$^{18}$O$(p,\alpha)^{15}$N reaction (for more details see
Cristallo et al.\,\cite{cri06}). However, even if models account for a substantial AGB production,
we confirm the previous claim by
Jorissen et al.\,(\cite{jo92}) that the theoretical predictions
underestimate the fluorine overabundance observed in evolved AGB
stars. A possible solution may be a more efficient production of
$^{15}$N in the He-rich inter-shell zone. Perhaps, an additional
and different source of fluorine should be sought. We have to note,
of course, that, as the fluorine abundance in the NGC 1846 stars has been derived
from only one line blend, a systematic error cannot be excluded.

%The low value
%of F shown by the less evolved stars in our sample is a
%consequence of the first dredge up.
%%%%%%%%%%

\subsection{Dredge up and pulsation properties}
\begin{figure}
\resizebox{\hsize}{!}{\includegraphics{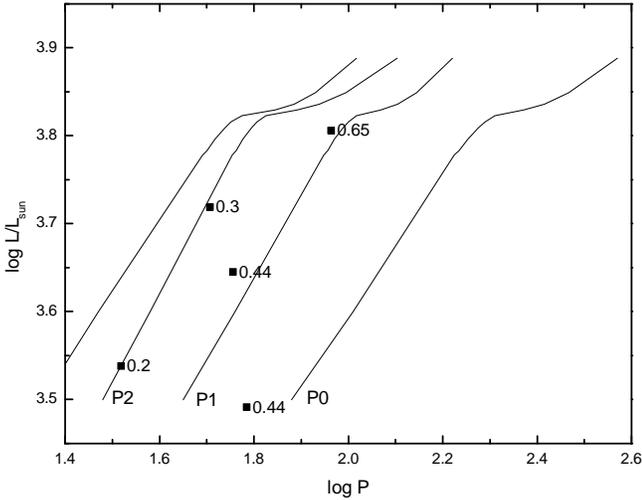}} \caption{The
log($L/L_{\sun}$) - log $P$ diagram for NGC 1846 adapted  from
Lebzelter \& Wood (\cite{LW07}). Pulsation models for fundamental
mode and the first three overtone modes are shown together with
five O-rich stars of our sample.} \label{pl}
\end{figure}

The detailed variability study on the AGB stars in NGC 1846
undertaken in Lebzelter \& Wood (\cite{LW07}) allows us to
investigate the observed star-to-star differences in the C/O ratio
in light of the pulsational properties of these stars. In
Fig.\,\ref{pl} we show the location of our sample stars in the
P-L-diagram of the cluster. Data are taken from Lebzelter \& Wood.
LE9 could not be plotted due to the lack of any detected
periodicity in its small-amplitude light variability. The other
stars can be easily attributed to a pulsation mode: LE13 and LE16
are first overtone pulsators, while LE8 and H39 pulsate in the
second overtone. Long secondary periods beyond the fundamental
mode period have been found in the light curves of several of these stars, but no
shorter secondary periods. As already pointed out in Lebzelter \& Wood
(\cite{LW07}), LE17 cannot be attributed unequivocally to any
pulsation sequence. 

\begin{figure}
\resizebox{\hsize}{!}{\includegraphics{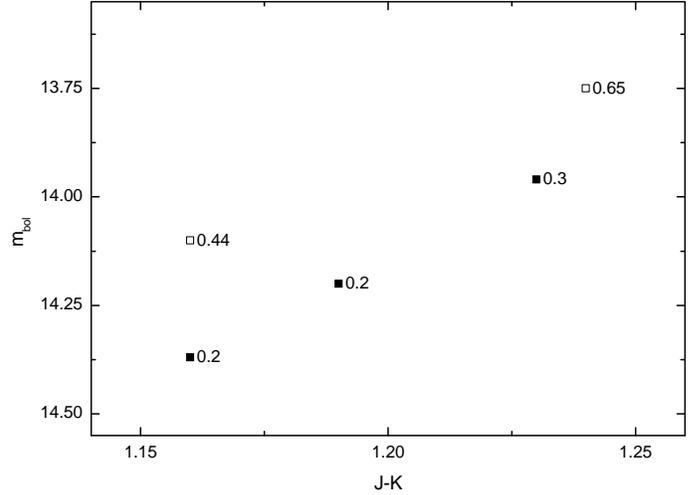}}
\caption{Location of the O-rich stars in a color-magnitude diagram
with the pulsation mode of each object indicated. Open boxes
denote first overtone pulsators, while filled boxes mark stars with
second overtone pulsation. The star LE17 has been excluded.} \label{cmd2}
\end{figure}

Stars located on the first overtone sequence show a higher C/O
ratio  than objects on the second overtone sequence
(Fig.\,\ref{pl}). As can be seen this cannot be attributed to a
simple difference in luminosity as the two groups clearly overlap.
This is further illustrated in Fig.\,\ref{cmd2} where we marked
the sample stars in the color-magnitude diagram according to their
pulsation mode. For
LE9 (C/O=0.2), no periodic change was found by Lebzelter \& Wood
(\cite{LW07}). Assuming that it would have been more likely to
find a first overtone pulsation than a shorter period higher
overtone pulsation, we attributed this star to the group of higher
overtone pulsators. These findings suggest that in addition to a
relation between the C/O ratio and the luminosity, seems to exist
some dependency of this ratio on the pulsation mode. At this point
it is not clear how to explain the existence of such a relation.
As mentioned above, all sample stars are on the AGB, we can exclude
an RGB contamination. From the difference in the C/O ratio, we can
assume that the sample stars are in at least three different
thermal pulse cycles. Both from our models and from the literature
(e.g.~Vassiliadis \& Wood \cite{VW93}), we would expect that the
stars become cooler from cycle to cycle. As the two C enhanced
stars LE16 and LE13 would form a sequence at higher temperatures
than the other stars, it can be excluded that a difference in the
thermal pulse cycle is responsible for the observed differences in
pulsation mode and temperature.

A difference in pulsation mode between two stars of the same
luminosity requires a difference in some other global
characteristic of the star. This could be a difference in mass.
Mackey \& Broby Nielsen (\cite{MB07}) report indications for two
stellar populations in NGC 1846 separated in age by about 300 Myr.
At a given time we may then find AGB stars of different masses at
the same time on the AGB. Indeed such an age or mass difference
could easily account for the observed color difference
(Fig.\,\ref{cmd2}). This would also be compatible with the
observed C/O ratios as the more massive stars may show a more
efficient dredge up. Furthermore, the starting abundances of the
two populations may be different, which could enhance the observed
difference in C/O and $^{12}$C/$^{13}$C.

It has to be stressed that, even in the case of a single stellar
population, a scatter in the abundances of C and O on the main
sequence and after the first dredge up may be present. Such an
abundance scatter has been found in various clusters.

\subsection{The S-type stars}
\begin{figure}
\resizebox{\hsize}{!}{\includegraphics{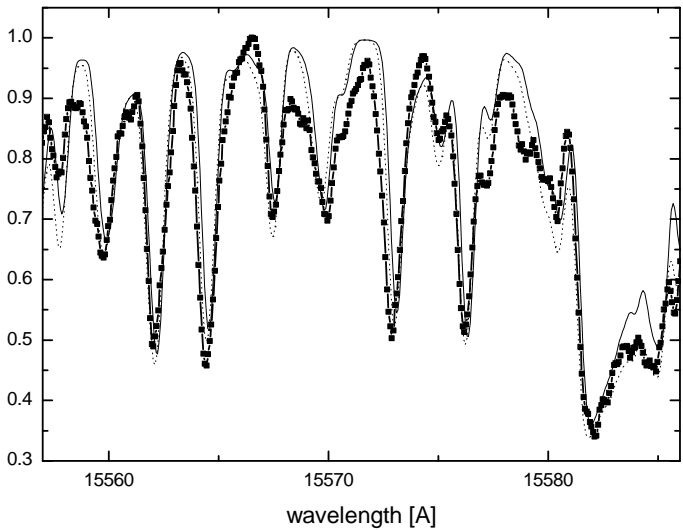}}
\caption{$H$ band spectrum of LE13 compared with models of 3600\,K, log\,g$=$0, C/O$=$0.6 (solid line) and 0.7 (dotted line),
respectively.}
\label{f:LE13}
\end{figure}

An interesting case is LE13. This star has been classified as
spectral type S3/3 by Lloyd Evans (\cite{LE83}). Indeed the ZrO
band head at 6345\,{\AA} is clearly visible in the spectrum
presented by him (see his Fig.\,1). According to its spectral
classification, TiO-bands and ZrO bands should be of similar
strength (Keenan \& Boeshaar \cite{KB80}). YO should occur, but is
not covered by the spectra shown by Lloyd Evans. 
As expected for a star with S-type spectral characteristics,
the C/O ratio is clearly enhanced in LE13 with a value between
0.6 and 0.7.
In Fig.\,\ref{f:LE13} we compare
the observed spectrum with our model for both cases (C/O=0.6 and C/O=0.7).
The model obviously fits the observations quite well.
As mentioned above, this model was calculated with an increased microturbulent velocity
to allow for a reasonable fit of the observed line profiles.
A change in the C/O ratio to 0.7 gives a similar fit, but a further increase in
the C/O ratio gets problematic as the CN lines, 
especially the one at 15567\,{\AA}, become too strong. Compensating for this with a
higher temperature would give a bad fit of
the OH line close to 15570\,{\AA} and too strong a CO band head. Similarly, changes
in log\,g cannot compensate for a C/O value close
to one. LE13 is thus an S-type star with C/O clearly below 1.

Lloyd Evans (\cite{LE83}) also classified a second star in this
cluster as of spectral type S3/1, namely LE8. LE8 has an even
lower C/O ratio of only about 0.3. Direct comparison with the
spectra of LE9 and H39 does indicate that the star has an
increased C/O ratio, but it is far below the expectations for an
S-star. Unfortunately, Lloyd Evans (\cite{LE83}) presented no visual
spectrum of this star.

A C/O ratio of only 0.65 for an S-type star is in agreement with the
oxygen overabundance in the Magellanic Cloud discussed above. If
this the case for NGC 1846, and the comparison of the other stars
with the expectations from the model suggests this, then we would
require more dredge up events to produce a C/O ratio close to one
than in the case of a solar abundance pattern. In other words, the
abundance of s-process enriched material on the surface may be
high enough to see e.g. ZrO bands, but the amount of C dredged up
does not yet equal the amount of O due to the high starting value
of the O abundance. Comparing the C/O ratio of LE13 and LE8, we
would expect to see less ZrO in LE8, which indeed agrees
with the spectral classes from Lloyd Evans (\cite{LE83}).

Piccirillo (\cite{picci}) pointed out that, only for stars with
T$<$3000\,K, would we expect that the occurrence of ZrO bands
correlates with a C/O close to one. For hotter stars, an
increased abundance of Zr is the main reason for a visibility of
ZrO in the spectrum. As the stars in our sample all have
temperatures above 3000\,K, the occurrence of ZrO bands is not
surprising. We think that a combination of this effect and an
enhancement of O in the pre-AGB composition of the star explains
the observed C/O ratio. In this context we also note that Smith \&
Lambert (\cite{SL90}) report a C/O for the two intrinsic
(i.e. AGB, see Jorissen et al. \cite{jorissen93}) S-stars NQ Pup
and V679 Oph of 0.29 and 0.75, respectively. Thus the C/O values
we found for the two S-type stars in NGC 1846 look
realistic.

\subsection{The C-stars}
The study of the C-stars in this cluster was hampered by these difficulties 
deriving basic parameters and abundances. Still a few points can be noticed. The C/O ratio
we find is rather high. There is quite a gap in our sample between the highest C/O ratio of an O-rich star
(LE13, 0.65) and the lowest value of a C-rich star (LE11, 1.7). There are still two candidates
in the cluster, LE12 and LW2 (see Lebzelter \& Wood \cite{LW07}), which may fill this gap,
but unfortunately we have no data for these stars. The value is also high compared to
the values found by Lambert et al.\,(\cite{lambert86}, see also Smith \& Lambert \cite{SL90}) for field stars, where the maximum
C/O is 1.6.  Such an occurrence is not surprising, however, since the oxygen content is lower in NGC 1846, even
considering [O/Fe]=0.2, when compared to field stars. 
From the abundance of HCN and C$_{2}$H$_{2}$, Matsuura et al.\,(\cite{Matsuura06}) find that the C/O
ratio in their sample of LMC carbon stars should be higher than 1.7, which would agree with
our findings.

Of particular interest is the carbon isotopic ratio derived for the 2 carbon stars in our sample.
 Although the C/O ratio
increases, the $^{12}$C/$^{13}$C remains close to the highest value found for the O-rich stars in
this cluster. Basing on the only 2 C stars in the sample, we can tentatively conclude that 
some saturation level exists around 60 for the carbon isotopic ratio.
In contrast to this figure, the evolutionary models predict a continuous increase in the $^{12}$C/$^{13}$C ratio,
 as due to the dredge up of primary $^{12}$C. Indeed,
 for C/O$>$1, our reference model predicts values definitely higher than 100.
Our finding confirms what has already been observed in galactic C-stars
(Smith \& Lambert \cite{SL90}, Abia et al. \cite{a01}), which shows a clustering between 50 and 70.

The most obvious explanation for this observational pattern
 is the occurrence of a mixing process able to bridge the radiative gap between the cool bottom of the convective envelope
and the hot H-burning zone. A similar process seems to explain some abundance anomalies in red giants
stars (see, e.g., Charbonnel \cite{ch95}). It is usually referred to as extra or deep mixing. More recently,
Nollett et al.~(\cite{no03}) have also proposed 
a deep circulation in low mass AGB stars as a solution to the puzzling isotopic anomalies
found in some pre-solar grains. They named this circulation the cool bottom process.
In spite of all the observational evidence, a
common consensus on the physical mechanism driving this mixing has not been established yet (e.g.
rotational induced instabilities, magnetically induced circulation, gravity wave, or thermohaline mixing -- see e.g. Busso \cite{Busso07}
for a recent review).

The scenario emerging from the evolutionary sequence of $^{12}$C/$^{13}$C versus C/O in NGC 1846 is that of a moderate
deep mixing, affecting the $^{13}$C surface abundance in the late part of the AGB only, when stars become C-rich.
Indeed, while the abundance derived from O-rich stars in NGC 1846 are all compatible with the prediction
of models with no extra mixing,
the $^{12}$C/$^{13}$C ratios we find for the two C stars indicate that their envelope material
could have been exposed to a temperature on the
order of 30-40\,$\cdot\,10^6$\,K. At that temperature and on an AGB timescale, $^{12}$C is partially converted into $^{13}$C,
but only marginally into $^{14}$N.

From the models, we find that, during the AGB the bottom of the convective envelope,
whose temperature never exceeds \mbox{$4\cdot 10^6$\,K},
moves closer to the H-burning zone. Just before the first thermal pulse, a region as large as
\mbox{$2.2\cdot 10^{-3}$~M$_\odot$} separates
the innermost convectively unstable layer from the shell where the temperature is \mbox{$40\cdot\,10^6$\,K}, but before the
10th TP, this region is reduced to \mbox{$4\cdot 10^{-4}$~M$_\odot$}. Such an occurrence suggests that the effect of
a possible extra mixing should be stronger in the late part of the AGB than at the beginning.

\begin{figure}
\resizebox{\hsize}{!}{\includegraphics{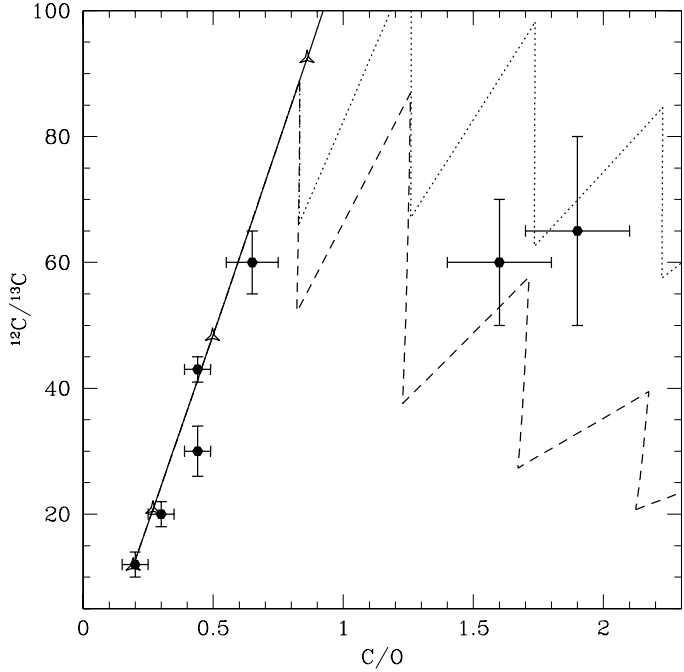}}
\caption{C/O versus $^{12}$C/$^{13}$C for all the sampled stars.
Predicted values from
models with and without extra-mixing are shown. Solid line refers to our
reference model, while the dotted and dashed ones refer to models with T$_{lim}$=35\,$\cdot 10^6$ K 
and T$_{lim}$=40\,$\cdot 10^6$ K, respectively (see text for details).}
\label{obsmeetmodels2}
\end{figure}

To simulate such a scenario, we calculated an additional model with the same input parameters as the
reference model, but artificially including an extra-mixing. We assume that such an extra-mixing is confined
 within a region
as large as $10^{-3}$~M$_\odot$ below the convective envelope, but only if the temperature is lower than a maximum value
(T$_{lim}$). The average convective velocity is fixed to 100 cm s$^{-1}$ \footnote{This is a slow mixing,
when compared to the typical
convective velocity in the envelope of an AGB star, which is of the order of $10^5$ cm/s. We have verified, however,
that the C/O and $^{12}$C/$^{13}$C ratios are only marginally affected by the choice of this parameter.}.
The most important parameter is, here, the value of the maximum temperature, as shown in Fig.\,\ref{obsmeetmodels2},
 where we reported the $^{12}$C/$^{13}$C versus C/O relations obtained for T$_{lim}$=35 and 40 ($10^6$ K).

This phenomenological model is useful for inferring the extension of the region affected by the supposed extra-mixing and
its maximum temperature, but a more physical description of the mechanism supporting this phenomenon is needed
to derive more quantitative results. Either way, we stress that the  $^{12}$C/$^{13}$C versus C/O sequences
for intermediate age clusters are the best way to understand and to constrain the deep mixing in AGB stars.

\section{Conclusions}
We derived C/O ratios and $^{12}$C/$^{13}$C
ratios for the first time  for a sample of AGB stars above the dredge-up mass limit
within one stellar cluster. In
this way we tested how the various third dredge-up events change
the surface abundances in a homogeneous sample of AGB stars.
Oxygen-rich stars with clear indications of a third dredge-up could
be identified from enhanced values of these two quantities, which
lie between 0.2 and 0.65 for C/O and 12 to 60 for
$^{12}$C/$^{13}$C. The cluster also contains a few
C-stars, but only rather uncertain values for the C/O ratio could
be derived. A more detailed analysis for two C-stars was possible
leading to C/O$=$1.7 and 1.9 and $^{12}$C/$^{13}$C$=$60 and 65.

As expected, the C/O ratio is correlated with the $^{12}$C/$^{13}$C
ratio.  We find good agreement with model predictions for both
ratios assuming an oxygen overabundance of +0.2 dex in the
cluster. The measured C/O ratios are in reasonable agreement with
the predicted interpulse values. This is the first direct test of
the models concerning dredge up before the stars become C-stars.

A possible correlation between C/O ratio and pulsation mode was
found.  The difference in pulsation mode may also help to explain
the observed scatter in C/O along the AGB of the cluster, although
no final understanding of this relation could be achieved. The
findings may be related to the existence of two populations within
the cluster with a small difference in age. Further investigations
of the populations in this cluster are needed.

An increase in the fluorine abundance along the AGB has been found.
This confirms nucleosynthesis in AGB stars as one of the origins of
this element.

Two stars in the cluster were classified as having spectral type
S in the past. We find a C/O of 0.65 and 0.3 for these
objects. While the C/O ratio is thus clearly less
than 1, we argue that the comparably high temperature of the stars
allows for S-type characteristics at this C/O.

The $^{12}$C/$^{13}$C ratio found in the C-stars cannot be explained with
standard models for the third dredge up. They indicate the need for a moderate
extra-mixing, affecting the late part of the AGB evolution.

\begin{acknowledgements}
This work was supported by the Austrian FWF under project number
P18171-N02. MTL has been supported by the Austrian Academy of Sciences
(DOC programme). Based on observations obtained at the Gemini Observatory, which is
operated by the Association of Universities for Research in Astronomy,
Inc., under a cooperative agreement with the NSF on behalf of the
Gemini partnership: the National Science Foundation (United States),
the Science and Technology Facilities Council (United Kingdom), the
National Research Council (Canada), CONICYT (Chile), the Australian
Research Council (Australia), CNPq (Brazil), and SECYT (Argentina).
The observations were obtained with the Phoenix infrared spectrograph,
which was developed and is operated by the National Optical Astronomy
Observatory.  The spectra were obtained as part of program
GS-2005B-C-7.
\end{acknowledgements}

{}
%---------------------------------------------------------------------------------------------------------------------


\begin{thebibliography}{}

\bibitem[2001]{a01} Abia, C., Busso, M., Gallino, R., Dom\'inguez, I., Straniero, O., Isern, J 2001,
\apj, 559, 1117
\bibitem[1989]{AG89} Anders, E., \& Grevesse, N. 1989, \gca, 53, 197
\bibitem[2000]{Ari2000} Aringer, B. 2000, PhD thesis, University of Vienna
\bibitem[2002]{Ari2002} Aringer, B., Kerschbaum, F., J{\o}rgensen, U.~G. 2002, \aap, 395, 915
%\bibitem[2002]{BHS02} Beasley, M.A., Hoyle, F., Sharples, R.M. 2002,
%\mnras, 336, 168
\bibitem[2007]{Busso07} Busso, M. 2007, ASPC 378, 26
\bibitem[1999]{Busso99} Busso, M., Gallino, R., Wasserburg, J. 1999, ARA\&A, 37, 239
\bibitem[2001]{Busso01} Busso, M., Gallino, R., Lambert, D.L., Travaglio, C., Smith, V.V. 2001,
\apj, 557, 802
\bibitem[1995]{ch95} Charbonnel, C. 1995, \apj, 453, 41
\bibitem[1998]{chi98} Chieffi, A., Limongi, M., Straniero, O. 1998 \apj, 502, 737
\bibitem[2006]{cri06} Cristallo, S., Gallino, R., Straniero, O.,
Piersanti, L. Dom\'inguez, I. 2006 \memsai, 77, 774.
\bibitem[2007]{cri07} Cristallo, S., Straniero, O., Lederer, M.T., Aringer,
B. 2007 \apj, 667, 489
\bibitem[2005]{cnlist05} Davis, S., Wallace, L., Brault, J., Engleman, R. 2005, The CN spectrum
from the infrared to the ultraviolet. NSO Technical Report
\bibitem[2003]{cunha03} Cunha, K., Smith, V.V., Lambert, D.L., Hinkle, K.H. 2003 \aj 126, 1305 
\bibitem[2006]{deni06} Denissenkov, P.A. Pinsonneault, M., Terndrup, D.M. \apj 651, 438
 \bibitem[1983]{FIO83} Freeman, K.C., Illingworth, G., Oemler, A. 1983, \apj, 272, 488
\bibitem[1990]{Frogel90} Frogel, J.A., Mould, J., Blanco, V.M. 1990, \apj, 352, 96
\bibitem[1995]{girardi95} Girardi, L., Chiosi, C., Bertelli, G., Bressan,A. 1995, \aap, 298, 87
\bibitem[2000]{gratton} Gratton, R.G., Sneden, C., Carretta, E., Bragaglia, A. 2000, \aap, 354, 169
\bibitem[1993]{GN93} Grevesse, N., Noels, A. 1993, Physica Scripta Volume T, 47, 133
\bibitem[2006]{grocho06} Grocholski, A.J., Cole, A.A., Sarajedini, A., et al.
2006, \aj, 132, 1630
\bibitem[1975]{MARCS75} Gustafsson, B., Bell, R.~A., Eriksson, K., Nordlund, A. 1975, \aap,
  42, 407
\bibitem[1987]{harris87} Harris, M.J., Lambert, D.L., Hinkle, K.H., Gustafsson, B., Eriksson, K.
1987, \apj, 316, 294
\bibitem[2000]{Herwig00} Herwig, F. 2000, \aap, 360, 952
%\bibitem[2005]{Herwig05} Herwig, F. 2005, \araa, 43, 435
\bibitem[2000]{Hill00} Hill, V., Fran{\c c}ois, P., Spite, M., Primas, F., Spite, F. 2000, \aap, 364, L19
\bibitem[1998]{phoenix98} Hinkle, K.H., Cuberly, R., Gaughan, N., et al. 1998, Proc. SPIE 3352, 810
\bibitem[2000a]{Houdi00a} Houdashelt, M.L., Bell, R.A., Sweigart, A.V. 2000a, \aj, 119, 1448
\bibitem[2000b]{Houdi00b} Houdashelt, M.L., Bell, R.A., Sweigart, A.V., Wing, R.F. 2000b, \aj, 119, 1424
\bibitem[1983]{ib83} Iben, I.Jr., Renzini, A. 1983 \araa, 21, 271.
\bibitem[1992]{MARCS92} J{\o}rgensen, U.~G., Johnson, H.~R., Nordlund, A. 1992, \aap, 261, 263
\bibitem[1992]{jo92} Jorissen, A., Smith, V.V., Lambert, D.L. 1992 \aap, 261 164
\bibitem[1993]{jorissen93} Jorissen, A., Frayer, D.T., Johnson, H.R., Mayor, M., Smith, V.V. 1993, \aap, 271, 463
\bibitem[2007]{KL07} Karakas, A., Lattanzio, J. 2007, PASA, 24, 103
\bibitem[1980]{KB80} Keenan, P.C., Boeshaar, P.C. 1980, \apjs, 43, 379
\bibitem[2000]{VALD} Kupka, F.~G., Ryabchikova, T.~A., Piskunov, N.~E., Stempels, H.~C., Weiss, W.~W. 2000, BaltA, 9, 590
\bibitem[1986]{lambert86} Lambert, D.L., Gustafsson, B., Eriksson, K., Hinkle, K.H. 1986, \apjs, 62, 373
\bibitem[2003]{LH03} Lebzelter, T., Hron, J. 2003, \aap, 411, 533
\bibitem[2007]{LW07} Lebzelter, T., Wood, P.R. 2007, \aap, 475, 643
\bibitem[2007]{LA07} Lederer, M.~T., Aringer, B. 2007, 	AIP Conf. Proc. submitted, arXiv:0712.2772v2
\bibitem[1980]{LE80} Lloyd Evans, T. 1980, \mnras, 193, 87
\bibitem[1983]{LE83} Lloyd Evans, T. 1983, \mnras, 204, 985
\bibitem[2003]{lo03} Lodders, L. 2003 \apjl, 591, 1220
 \bibitem[2001]{Loidl01} Loidl, R., Lan{\c c}on, A., J{\o}rgensen, U.~G. 2001, \aap, 371, 1065
\bibitem[2007]{MB07} Mackey, A.D., Broby Nielsen, P. 2007, \mnras, 379, 151
\bibitem[1999]{Marigo99} Marigo, P., Girardi, L., Bressan, A. 1999, \aap, 344, 123
\bibitem[2006]{Matsuura06} Matsuura, M., Wood, P.R., Sloan, G.C., et al. 2006, \mnras, 371, 415
\bibitem[2003]{no03}Nollett, K.M., Busso, M., Wasserburg, G.J. 2003 \apj, 582, 1036
\bibitem[1991]{OSSH91} Olszewski, E.W., Schommer, R.A., Suntzeff, N.,
Harris, H.C. 1991, \aj, 101, 515
\bibitem[1980]{picci} Piccirillo, J. 1980, \mnras, 190, 441
\bibitem[2007]{pier07} Piersanti, L., Straniero, O., Cristallo, S. 2007, \aap, 462, 1051
\bibitem[1992]{SOSH92} Schommer, R.A., Olszewski, E.W., Suntzeff, N., Harris, H.C.
1992, \aj, 103, 447
\bibitem[2007]{SCS07} Schuler, S.C., Cunha, K., Smith, V.V., et al. 2007, \apj, 667, L81
\bibitem[1990]{SL90} Smith, V.V., Lambert, D.L. 1990, \apjs, 72, 387
\bibitem[1991]{sneden} Sneden, C. 1991, in IAU Symp. 145, Evolution of Stars: The Photospheric
Abundance Connection, ed. G. Michaud \& A. Tutukov (Dordrecht: Kluwer),
235
\bibitem[2004]{stancliffe04} Stancliffe, R.J., Izzard, R.G., Tout, C.A. 2004, \mnras, 356, 4
\bibitem[1997]{stra97} Straniero, O., Chieffi, A., Limongi, M., Busso, M., Gallino, R., Arlandini, C. 1997
\apj, 478, 332
\bibitem[2003]{SDCG03} Straniero, O., Domínguez, I., Cristallo, S., Gallino, R. 2003, PASA, 20, 389
\bibitem[2006]{stra06} Straniero, O., Gallino, R., Cristallo, S. 2006
\nphysa 777, 311
 \bibitem[1998]{T98} Tanab\'{e}, T., Nishida, S., Nakada, Y., et al. 1997,
\apss, 255, 407
\bibitem[2001]{vaneck01} van Eck, S., Goriely, S., Jorissen, A., Plez, B. 2001, Nature, 412, 793
\bibitem[1993]{VW93} Vassiliadis, E., Wood, P.R. 1993, \apj, 413, 641
\bibitem[1997]{Windsteig97} Windsteig, W., Dorfi, E.~A., H{\"o}fner, S., Hron, J., Kerschbaum, F. 1997, \aap, 324, 617
\bibitem[2004]{Wood04} Wood, P.R., Olivier, E.A., Kawaler, S.D., 2004, \apj, 604, 800
\bibitem[2005]{zhang05} Zhang, Y., Liu, X.-W. 2005 \apj 631, 61
\end{thebibliography}
\end{document}